\documentclass[envcountsame]{llncs}

\usepackage{amsmath}
\usepackage{amsfonts}
\usepackage{amssymb}
\usepackage{latexsym}
\usepackage{stmaryrd}
\usepackage{array}
\usepackage{exscale}
\newcommand{\nc}{\newcommand}
\newcommand{\ol}{\overline}

\newcommand{\es}{\emptyset}
\newcommand{\sm}{\setminus}
\newcommand{\ve}{\varepsilon}
\newcommand{\vp}{\varphi}

\newcommand{\bc}{\bigcup}

\newcommand{\Lra}{\Leftrightarrow}

\newcommand{\ra}{\rightarrow}

\newcommand{\lra}{\leftrightarrow}

\newcommand{\sse}{\subseteq}

\newcommand{\spe}{\supseteq}
\newcommand{\fa}{\forall}
\newcommand{\ex}{\exists}
\newcommand{\mr}{\mathrm}
\newcommand{\mc}{\mathcal}
\newcommand{\mf}{\mathfrak}

\newcommand{\DMO}{\DeclareMathOperator}
\newcommand{\DST}{\displaystyle}

\newcommand{\NN}{\mathbb{N}}
\newcommand{\NNZ}{\NN_0}
\newcommand{\ZZ}{\mathbb{Z}}

\newcommand{\RR}{\mathbb{R}}

\newcommand{\PP}{\mathbb{P}}

\mathchardef\breakingcomma\mathcode`\,
{\catcode`,=\active
  \gdef,{\breakingcomma\discretionary{}{}{}}
}

\newcommand{\mb}{{\:|\:}} 
\newcommand{\set}[1]{\{ #1 \}}
\newcommand{\setb}[1]{\big \{ \, #1 \, \big \}}
\nc{\simlvi}[1]{\!\sim_{#1}}
\nc{\apprel}[3]{{#1}(#2)_{(#3)}} 
\nc{\cmpli}[1]{\complement^1_{#1}} 
\nc{\cmplzi}[1]{\complement^0_{#1}} 
\nc{\cmplzoi}[1]{\complement^*_{#1}} 

\nc{\zf}{\mr{ZF}}
\nc{\zfmf}{\zf^0} 
\nc{\zfc}{\mr{ZFC}}
\nc{\zfcmf}{\zfc^0} 
\nc{\bst}{\mr{BST}} 
\newcommand{\tb}[2]{\set{#1, \dots, #2}} 
\providecommand{\abs}[1]{\lvert #1 \rvert} 
\providecommand{\norm}[1]{\lVert #1 \rVert} 
\makeatletter
\DeclareRobustCommand{\genericinterval}[2]{%
  \@ifstar{\genericinterval@star{#1}{#2}}{\genericinterval@nostar{#1}{#2}}}
\newcommand{\genericinterval@star}[4]{\mathopen{}\mathclose{\left#1#3,#4\right#2}}
\newcommand{\genericinterval@nostar}[4]{\mathopen{#1}#3,#4\mathclose{#2}}

\makeatother
\nc{\untit}[2]{{#1}^{#2 \downarrow}} 
\nc{\obit}[2]{{#1}^{#2 \uparrow}} 
\nc{\inzEKi}[1]{\mc{I}^{\mr{V}}_{#1}}

\nc{\inzKEi}[1]{\mc{I}^{\mr{E}}_{#1}}

\nc{\adjEi}[1]{\mc{A}^{\mr{V}}_{#1}}

\nc{\BD}[1]{{#1}\text{-}\mr{BD}}

\DeclareMathOperator{\pot}{\PP} 
\DeclareMathOperator{\pote}{\PP_f} 
\nc{\konv}[2]{{#1}[{#2}]} 
\nc{\actpres}[1]{\Phi_{#1}} 
\newcommand{\ceil}[1]{\lceil #1 \rceil}
\nc{\Prim}{\mc{PR}} 

\nc{\sselr}{\sse^{\mapsto}}
\nc{\sserl}{\sse^{\mapsfrom}}
\nc{\spelr}{\spe^{\mapsto}}
\nc{\sperl}{\spe^{\mapsfrom}}
\nc{\ball}[1]{\mr{B}^{#1}} 
\nc{\oball}[1]{\breve{\mr{B}}^{#1}} 
\nc{\pball}[1]{\dot{\mr{B}}^{#1}} 
\nc{\prr}[1]{\dot{\RR}^{#1}} 
\nc{\sph}[1]{\mr{S}^{#1}} 
\nc{\ssim}[1]{s\sigma_{#1}} 
\nc{\koerper}[1]{\norm{#1}}
\nc{\Ccovdim}{\mc{CD}}
\nc{\Cinddim}{\mc{SID}}

\nc{\CInddim}{\mc{LID}}

\DeclareMathOperator{\diffop}{D} 
\DeclareMathOperator*{\diffoplimit}{D} 
\nc{\diffopc}[1]{\sideset{_{#1}}{}\diffoplimit} 
\nc{\diffopp}[1]{\diffop_{#1}} 
\nc{\diffopcp}[2]{\sideset{_{#2}}{_{#1}}\diffoplimit} 
\nc{\meanH}[2]{\mf{M}_{#1,#2}} 
\nc{\emean}[2]{\mf{M}_{\exp_{#1},#2}} 
\DeclareMathOperator{\mor}{Mor}
\DeclareMathOperator{\Hom}{Hom} 
\nc{\autoerw}[1]{\mr{Aut}^{#1}} 
\nc{\komma}[2]{(#1 \downarrow #2)} 
\nc{\Kmat}{\mf{MAT}} 
\nc{\Khmat}{\mf{HMAT}} 
\nc{\homfun}[1]{\mor_{#1}(-_1,-_2)} 
\nc{\homfunae}[1]{\mor_{#1}(-_1)} 
\nc{\homfunbe}[1]{\mor_{#1}(-_2)} 
\nc{\homfunxy}[3]{\mor_{#1}(#2(-_1), #3(-_2))}
\nc{\homfunx}[2]{\mor_{#1}(#2(-_1), -_2)}
\nc{\homfuny}[2]{\mor_{#1}(-_1, #2(-_2))}
\nc{\homfuna}[2]{\mor_{#1}(#2, -)} 
\nc{\homfunb}[2]{\mor_{#1}(-, #2)} 
\nc{\hhomfuna}[2]{\Hom_{#1}(#2, -)} 
\nc{\hhomfunb}[2]{\Hom_{#1}(-, #2)} 
\newcommand{\Va}{\mc{V\hspace{-0.1em}A}}

\newcommand{\Lit}{\mc{LIT}}

\newcommand{\Cls}{\mc{CLS}}

\newcommand{\Pcls}[1]{#1\mbox{--}\Cls}

\newcommand{\Pass}{\mc{P\hspace{-0.32em}ASS}}
\newcommand{\epa}{\pab{}} 
\newcommand{\Tass}{\mc{T\hspace{-0.35em}ASS}}
\newcommand{\Sat}{\mc{SAT}}

\newcommand{\Usat}{\mc{USAT}}

\newcommand{\Musat}{\mc{M\hspace{0.8pt}U}} 
\newcommand{\Musati}[1]{\Musat_{\!#1}} 
\newcommand{\Smusat}{\mc{S}\Musat} 
\newcommand{\Smusati}[1]{\Smusat_{\!#1}}

\nc{\Clsoo}{\Cls^{1,1}} 
\DeclareMathOperator{\lit}{lit}
\DeclareMathOperator{\var}{var}

\DMO{\dos}{ds} 
\DMO{\mdos}{mds} 
\newcommand{\Clash}{\mc{HIT}} 

\newcommand{\Uclash}{\mc{U}\Clash} 
\newcommand{\Uclashi}[1]{\Uclash_{\!\!#1}}

\newcommand{\Lean}{\mc{LEAN}}

\DMO{\premr}{ax} 
\DMO{\concr}{C} 
\DMO{\allcr}{cl} 

\DMO{\thardness}{thd} 
\DMO{\phardness}{phd} 
\DMO{\whardness}{awid} 
\DMO{\dep}{dep} 
\DMO{\hts}{hs} 
\DMO{\semspace}{css} 
\DMO{\resspace}{crs} 
\DMO{\treespace}{cts} 
\newcommand{\pab}[1]{\langle #1 \rangle}

\nc{\bth}[1]{\langle{#1}\rangle} 
\DeclareMathOperator{\aut}{Auk} 
\DeclareMathOperator{\autf}{Auk^r} 
\DeclareMathOperator{\autmax}{Auk\!\!\uparrow} 
\DeclareMathOperator{\nv}{N} 
\DeclareMathOperator{\na}{\nv_a} 

\DMO{\rsub}{r_S} 
\DMO{\rk}{r} 
\DMO{\ro}{\rk_1} 
\DMO{\rki}{\rk_{\infty}} 
\DMO{\rpl}{r^{pl}} 
\DMO{\ropl}{\rk_1^{pl}} 
\nc{\rslur}{\xrightarrow{\text{SLUR}}} 
\nc{\rslurs}{\rslur_{\!*}} 
\DMO{\slur}{slur} 
\nc{\Slur}{\mc{SLUR}} 
\nc{\rkslur}[1]{\xrightarrow{\text{SLUR}_{#1}}} 
\nc{\rkslurs}[1]{\rkslur{#1}_{\!*}} 
\nc{\Altsluri}[1]{\Slur(#1)}
\nc{\Altslurstari}[1]{\Slur\text{\textasteriskcentered}(#1)}
\nc{\Canoni}[1]{\mr{CANON}(#1)}
\nc{\rkslurstar}[1]{\xrightarrow{\text{SLUR\textasteriskcentered}#1}} 
\nc{\rkslursstar}[1]{\rkslurstar{#1}_{\!*}} 
\DMO{\slurstar}{\slur\!\text{\textasteriskcentered}}
\nc{\Urefc}{\mc{UC}}
\nc{\Propc}{\mc{PC}}
\nc{\Wrefc}{\mc{WC}} 
\DeclareMathOperator{\vdeg}{vd} 
\DeclareMathOperator{\minvdeg}{\mu\!\vdeg} 
\DMO{\varmvd}{\var_{\minvdeg}} 
\DMO{\nfc}{fc} 
\DMO{\maxnfc}{\nu\!\nfc} 
\nc{\svbf}{\mc{VB}} 
\nc{\svbfs}{\mc{VB}^*} 
\DMO{\potp}{pp} 
\DMO{\potprec}{NM} 
\DMO{\minnonmer}{VDM} 
\DMO{\minnonmerh}{VDH} 
\DMO{\maxsmar}{FCM} 
\DMO{\maxsmarh}{FCH} 
\DMO{\varsing}{\var_s} 
\DMO{\varosing}{\var_{1s}} 
\DMO{\varnosing}{\var_{\neg1s}} 
\nc{\Musatns}{\Musat'} 
\nc{\Musatnsi}[1]{\Musati{#1}'}
\nc{\Smusatns}{\Smusat'} 
\nc{\Smusatnsi}[1]{\Smusati{#1}'}
\nc{\Uclashns}{\Uclash'} 
\nc{\Uclashnsi}[1]{\Uclashi{#1}'}
\nc{\tsdp}{\xrightarrow{\text{sDP}}}
\nc{\tsdps}{\tsdp_{\!*}}
\nc{\tosdp}{\xrightarrow{\text{1sDP}}}
\nc{\tosdps}{\tosdp_{\!*}}
\DMO{\sdp}{sDP} 
\DMO{\osdp}{sDP_1} 
\nc{\cflmusat}{\mc{CF}\Musat} 
\nc{\cflmusati}[1]{\mc{CF}\Musati{#1}}
\nc{\cflimusat}{\mc{CFI}\Musat} 
\DMO{\sNF}{sNF} 
\DMO{\eqp}{eqp} 
\DMO{\sgp}{sp} 
\DMO{\singind}{si} 
\DMO{\osingind}{si_1} 
\DMO{\shyp}{svh} 
\DMO{\sdph}{ssh} 
\DMO{\msdph}{mss} 
\DMO{\osdph}{ssh_1} 
\DMO{\mosdph}{mss_1} 
\DMO{\mps}{mps} 
\DMO{\purec}{puc} 
\DMO{\doping}{D}
\nc{\glue}[4]{\operatorname{glue}((#1,#2), (#3,#4))} 
\nc{\gluea}[3]{#1 \mathbin{\boxplus}_{#3} #2} 
\DMO{\frl}{fl} 
\nc{\Con}{\mr{Con}}
\nc{\Log}{\mr{Log}}
\nc{\Lin}{\mr{Lin}}
\nc{\Pol}{\mr{Pol}}
\nc{\ExL}{\mr{ExL}}
\nc{\ExP}{\mr{ExP}}
\nc{\CTime}{\mr{CTime}}
\nc{\CSpace}{\mr{CSpace}}
\nc{\LTime}{\mr{LTime}}
\nc{\LSpace}{\mr{L}}
\nc{\NLSpace}{\mr{NL}}
\nc{\LinTime}{\mr{LinTime}}
\nc{\LinSpace}{\mr{LinSpace}}
\nc{\PTime}{\mr{P}}
\nc{\PSpace}{\mr{PSpace}}
\nc{\Np}{\mr{NP}}
\nc{\Conp}{\text{coNP}}
\nc{\NPSpace}{\mr{NPSpace}}
\nc{\CoNPSpace}{\mr{coNPSpace}}
\nc{\ELTime}{\mr{ELTime}}
\nc{\ELSpace}{\mr{ELSpace}}
\nc{\EPTime}{\mr{EPTime}}
\nc{\EPSpace}{\mr{EPSpace}}
\nc{\NEPTime}{\mr{NEPTime}}
\nc{\polydelta}[1]{\Delta_{#1}^{\mr P}}
\nc{\polypi}[1]{\Pi_{#1}^{\mr P}}
\nc{\polysigma}[1]{\Sigma_{#1}^{\mr P}}
\nc{\Ph}{\mr{PH}}

\nc{\Dp}{D^P}
\nc{\PllC}[2]{{\text{$\mr{PT}$/$\mr{WK}$}(#1, #2)}} 
\nc{\Nc}{\mr{NC}}
\nc{\Nci}[1]{\Nc^{#1}}
\nc{\Ac}{\mr{AC}}
\nc{\Aci}[1]{\Ac^{#1}}
\nc{\pmodpoly}{P / \mathrm{poly}}
\nc{\Wh}[1]{\mr{W}[#1]} 
\nc{\Rl}{\mr{RL}}
\nc{\coRl}{\mr{coRL}}
\nc{\Rp}{\mr{RP}}
\nc{\coRp}{\mr{coRP}}
\nc{\Zpp}{\mr{ZPP}}
\nc{\Bpp}{\mr{BPP}}
\nc{\Pp}{\mr{PP}}
\nc{\Reach}{\mr{STCON}} 
\nc{\Undreach}{\mr{USTCON}} 
\nc{\Pcol}[2]{\mr{COL}(#1,#2)} 
\nc{\Pscol}[2]{\mr{SCOL}(#1,#2)} 
\nc{\Psorcol}[2]{\mr{SORCOL}(#1,#2)} 
\DMO{\slp}{slp}
\nc{\Mss}{\mr{MSS}}
\nc{\Key}{\mr{KEY}}
\nc{\Keyi}[1]{\Key_{\!#1}}
\nc{\Nbmss}{N_{\mr{bm}}} 
\nc{\Nbkey}{N_{\mr{bk}}} 
\nc{\Rnb}{N_{\mr{b}}}
\nc{\Rnk}{N_{\mr{k}}}
\nc{\Rnr}{N_{\mr{r}}}

\nc{\Byte}{\mr{B}[8]}
\nc{\QByte}{\mr{B}[4,8]}
\nc{\KByte}{\mc{B}} 
\nc{\RQByte}{\mc{QB}} 

\nc{\ramz}[3]{\mr{ram}_{#1}^{#2}(#3)} 
\nc{\waez}[2]{\mr{vdw}_{#1}(#2)} 
\nc{\gtz}[2]{\mr{grt}_{#1}(#2)} 
\nc{\pdwaez}[2]{\mr{vdw}_{#1}^{\mr{pd}}(#2)} 
\nc{\absfeh}[1]{\delta_{#1}} 
\nc{\relfeh}[1]{\ve_{#1}} 
\usepackage{theorem} 
\usepackage{hyperref} 
\theorembodyfont{\rmfamily}

\theorembodyfont{}
\newenvironment{prf}{\noindent\textbf{Proof:}\;}{\par\noindent\ignorespacesafterend}

\newcommand{\Qed}{\hfill $\square$}

\nc{\bm}{\boldmath}
\nc{\bmm}[1]{\mbox{\bm$\DST #1$}}
\nc{\mi}[1]{\bmm{\mathrm{(#1):}} \quad}
\usepackage{enumerate}
\usepackage[all]{xy}

\newcommand{\Schrift}{paper}

\DeclareMathOperator{\nva}{\mathit{n}_A}
\DeclareMathOperator{\nvl}{\mathit{n}_L}

\newcommand{\oracle}{\mc{O}}
\newcommand{\oraclez}{\mc{O}_0}
\newcommand{\oracleo}{\mc{O}_1}
\newcommand{\oraclezo}{\mc{O}_{01}}
\newcommand{\algobs}{\mc{A}_{\mathrm{bs}}}
\newcommand{\algoz}{\mc{A}_0}
\newcommand{\algoo}{\mc{A}_1}
\newcommand{\schemezo}{\mc{S}_{01}}
\newcommand{\algozo}{\mc{A}_{01}}

\begin{document}

\pagestyle{headings} 

\title{Computing maximal autarkies with\\few and simple oracle queries}

\author{Oliver Kullmann\inst{1} and Jo{\~{a}}o Marques-Silva\inst{2}\thanks{This work is partially supported by SFI PI grant BEACON (09/IN.1/I2618), FCT grant POLARIS (PTDC/EIA-CCO/123051/2010) and national funds through FCT with reference UID/CEC/50021/2013.}}
\institute{
  Computer Science Department, Swansea University, UK\\
  \and
  INESC-ID, IST, University of Lisbon, Portugal
}

\maketitle

\begin{abstract}
  We consider the algorithmic task of computing a \emph{maximal autarky} for a clause-set $F$, i.e., a partial assignment which satisfies every clause of $F$ it touches, and where this property is destroyed by adding any non-empty set of further assignments. We employ SAT solvers as oracles here, and we are especially concerned with minimising the number of oracle calls. Using the standard SAT oracle, $\log_2(n(F))$ oracle calls suffice, where $n(F)$ is the number of variables, but the drawback is that (translated) cardinality constraints are employed, which makes this approach less efficient in practice. Using an extended SAT oracle, motivated by the capabilities of modern SAT solvers, we show how to compute maximal autarkies with $2 \sqrt{n(F)}$ simpler oracle calls, by a novel algorithm, which combines the previous two main approaches, based on the autarky-resolution duality and on SAT translations.
\end{abstract}

\section{Introduction}
\label{sec:intro}

A well-known application area of SAT solvers is the analysis of
over-constrained systems, i.e.\ systems of constraints that are
inconsistent.
A number of computational problems can be related with the analysis of
over-constrained systems. These include minimal explanations of
inconsistency, and minimal relaxations to achieve consistency.
Pervasive to these computational problems is the problem of computing
a ``maximal autarky'' of a propositional formula, since clauses satisfied by an
autarky cannot be included in minimal explanations of inconsistency or
minimal relaxations to achieve consistency. In the experimental study \cite{SIMML2014EfficientAutarkies} it was realised that using as few SAT calls as possible, via cardinality-constraints, performs much worse than using a linear number of calls. To use only a sublinear number of calls, without using cardinality constraints, is the goal of this \Schrift.

Given a satisfiable clause-set $F$ and a partial assignment $\vp$, in general $\vp * F$, the result of the application (instantiation) of $\vp$ to $F$, might be unsatisfiable. $\vp$ is an \emph{autarky} for (arbitrary) $F$ iff every clause $C$ of $F$ touched by $\vp$ (i.e., $\var(C) \cap \var(\vp) \ne \es$) is satisfied by $\vp$ (i.e., $\ex\, x \in C : \vp(x) = 1$). Now if $F$ is satisfiable, then also $\vp * F$ is satisfiable, since due to the autarky property holds $\vp * F = \set{C \in F : \var(C) \cap \var(\vp) = \es} \sse F$. Thus ``autarky reduction'' $F \leadsto \vp * F$ can take place (satisfiability-equivalently). An early use of autarkies is \cite{EIS76}, for the solution of 2-SAT. The notion ``autarky'' was introduced in \cite{MoSp85} for faster $k$-SAT decision, which can be seen as an extension of \cite{EIS76}. For an overview of such uses of autarkies for SAT solving see \cite{HvM09HBSAT}. Besides such incomplete usage (using only autarkies ``at hand''), the complete search for ``all'' autarkies (or the ``strongest'' one) is of interest. Either with (clever) exponential-time algorithms, or for special classes of clause-sets, where polynomial-time is possible, or considering only restricted forms of autarkies to enable polynomial-time handling; see \cite{Kullmann2007HandbuchMU} for an overview. In \cite{Kullmann2007ClausalFormZI,Kullmann2007ClausalFormZII} autarky theory is generalised to non-boolean clause-sets.

Finitely many autarkies can be composed to yield another autarky, which satisfies precisely the clauses satisfied by (at least) one of them; this was first observed in \cite{Ok98}. So complete autarky reduction for a clause-set $F$, elimination of clauses satisfied by some autarky as long as possible, yields a unique sub-clause-set, called the \emph{lean kernel} $\na(F) \sse F$, as introduced in \cite{Ku98e} and further studied in \cite{Ku00f}; we note that $F \in \Sat \Lra \na(F) = \top$, where $\top$ is the empty clause-set. Clause-sets without non-trivial autarkies are called \emph{lean}, and are characterised by $\na(F) = F$; the set of all lean clause-sets is called $\Lean$, and was shown to be coNP-complete in \cite{Ku00f}. A \emph{maximal autarky} for $F$ is one which can not be extended; note that a maximal autarky $\vp$ always exist, where $\vp = \epa$, the empty partial assignment, iff $F$ is lean. An autarky $\vp$ is maximal iff $\var(\vp) = \var(F) \sm \var(\na(F))$. Thus $\var(F) \sm \var(\na(F))$ is called the \emph{largest autarky var-set}. For a maximal autarky $\vp$ the result of the autarky reduction is $\na(F)$, while any autarky which yields $\na(F)$ is called \emph{quasi-maximal}.

\paragraph{Algorithmic problems associated with autarkies.}

The basic algorithmic problems related to general ``autarky systems'', which allow to specialise the notion of autarky, for example in order to enable polynomial-time computations, are discussed in \cite[Section 11.11.6]{Kullmann2007HandbuchMU}. Regarding \emph{decision problems}, for this \Schrift{} only one problem is relevant here, namely \texttt{AUTARKY EXISTENCE}, deciding whether a clause-set $F$ has a non-trivial autarky; the negation is \texttt{LEAN}, deciding whether $F \in \Lean$. An early oracle-result is \cite[Lemma 8.6]{Ku00f}, which shows, given an oracle for \texttt{LEAN}, how to compute \texttt{LEAN KERNEL} with at most $n(F)$ oracle calls (for all ``normal autarky systems'', using the terminology from \cite[Section 11.11]{Kullmann2007HandbuchMU}). We are concerned in this \Schrift{} with the \emph{functional problems}, where the four relevant problems are as follows, also stating the effort for checking a solution:

\texttt{NON-TRIVIAL AUTARKY}: Find some non-trivial autarky (if it exists; otherwise return the empty autarky). Checking an autarky is in $P$.

\texttt{QUASI-MAXIMAL AUTARKY} or \texttt{MAXIMAL AUTARKY}: Find a (quasi-)maximal autarky; by a trivial computation, from a quasi-maximal autarky we can compute a maximal one. Checking that $\vp$ is a quasi-maximal autarky for $F$ means checking that $\vp$ is an autarky (easy), and that $\vp * F$ is lean, and so checking is in coNP. A quasi-maximal autarky can be computed by repeated calls to \texttt{NON-TRIVIAL AUTARKY} (until no non-trivial autarky exists anymore).

\texttt{NON-TRIVIAL VAR-AUTARKY}: Find the var(iable)-set of some non-trivial autarky (if it exists; otherwise return the empty set). Checking that $V$ is the variable-set of an autarky means checking that $F[V]$, the restriction of $F$ to $V$, is satisfiable, thus checking is in NP.

\texttt{(QUASI-)MAXIMAL VAR-AUTARKY} or \texttt{LEAN KERNEL}: Compute the largest autarky var-set (or a quasi-maximal one), or compute the lean kernel; all three tasks are equivalent by trivial computations. Checking that $V$ is the largest autarky var-set means checking that $F[V]$ is satisfiable and that $\set{C \in F : \var(F) \cap V = \es}$ is lean, so checking is in $D^P$ (\cite{PW88}). The solution to \texttt{MAXIMAL VAR-AUTARKY} or to \texttt{LEAN KERNEL} is unique and always exists. The var-set of a quasi-maximal autarky can be computed by repeated calls to \texttt{NON-TRIVIAL VAR-AUTARKY}.

Just having the var-set of the autarky $\vp$ enables us to perform the autarky reduction $F \leadsto \vp * F$, namely $\vp * F = \set{C \in F : \var(C) \cap \var(\vp) = \es}$, but from the var-set $\var(\vp)$ in general we can not derive the autarky $\vp$ itself, which is needed to provide a certificate for the autarky-property. For example, $F$ is satisfiable iff $\var(F)$ is the largest autarky var-set, and in general without further hard work it is not possible to obtain the satisfying assignment from (just) the knowledge that $F$ is satisfiable. An interesting case is discussed in \cite[Subsection 4.3]{KullmannZhao2011Bounds} and (in greater depth) in \cite[Section 10]{KullmannZhao2010Extremal}, where we can compute a certain autarky reduction in polynomial-time, but it is not known how to find the autarky (efficiently). So \texttt{NON-TRIVIAL VAR-AUTARKY} is weaker than \texttt{NON-TRIVIAL AUTARKY}, and \texttt{MAXIMAL VAR-AUTARKY} is weaker than \texttt{MAXIMAL AUTARKY}. We tackle in this \Schrift{} the hardest problem, \texttt{MAXIMAL AUTARKY}.

To obtain a complexity calibration, we can consider the computational model where polynomial-time computation and (only) one oracle call is used. Then \texttt{MAXIMAL VAR-AUTARKY} is equivalent to \texttt{PARALLEL SAT}, which has as input a list $F_1, \dots, F_m$ of clause-sets, and as output $m$ bits deciding satisfiability of the inputs: On the one hand, given these $F_1, \dots, F_m$, make them variable-disjoint and input their union to the \texttt{MAXIMAL VAR-AUTARKY} oracle --- $F_i$ is satisfiable iff $\var(F_i)$ is contained in the largest autarky var-set. On the other hand it is an easy exercise to see, that for example via the translation $F \leadsto t(F)$ used in this \Schrift, introduced as $\Gamma_2$ in \cite{SIMML2014EfficientAutarkies}, we can compute the largest autarky var-set by inputting $t(F) \cup \set{\set{v_1}}, \dots, t(F) \cup \set{\set{v_n}}$ to \texttt{PARALLEL-SAT}, where $\var(F) = \set{v_1, \dots, v_n}$. Similarly it is easy to see that \texttt{MAXIMAL AUTARKY} is equivalent to \texttt{PARALLEL FSAT} (here now also the satisfying assignments are computed).

\paragraph{General approaches for the lean kernel.}

See \cite[Section 11.10]{Kullmann2007HandbuchMU} for an overview. A fundamental method for computing a (quasi-)maximal autarky, strengthened in this \Schrift, uses the \emph{autarky-resolution duality} (\cite[Theorem 3.16]{Ku98e}): the variables in the largest autarky var-set are precisely the variables not usable in any resolution refutation. The basic algorithm, reviewed as algorithm $\algoz$ in Definition \ref {def:A0} in this \Schrift{} (with a refined analysis), was first given in \cite{Ku01a} and somewhat generalised in \cite[Theorem 11.10.1]{Kullmann2007HandbuchMU}; see \cite{KullmannLynceSilva2005Autarkies} for a discussion and some experimental results. A central concept is, what in this \Schrift{} we call an \emph{extended SAT oracle} $\oraclezo$, which for a satisfiable input outputs a satisfying assignment, while $\oraclezo$ on an unsatisfiable input outputs the variables used by some resolution refutation. In order to also accommodate polynomial-time results, the oracle $\oraclezo$ may get its inputs from a class $\mc{C}$ of clause-sets, which is stable (closed) under removal of variables. However, for the new algorithm of this \Schrift{} (Algorithm $\algozo$ presented in Theorem \ref{thm:main}), we do not consider classes $\mc{C}$ as for $\algoz$, since the input is first transformed, and then also some clauses are added, which would complicate the requirements on $\mc{C}$. The other main method to compute autarkies uses reduction to SAT problems, denoted by $F \leadsto t(F)$ in this \Schrift, where the solutions of $t(F)$ correspond to the autarkies of $F$. This was started by \cite{LiffitonSakallah2008Trimming}, and further extended first in \cite[Subsection 11.10.4]{Kullmann2007HandbuchMU}, and then in \cite{SIMML2014EfficientAutarkies}, which contains a thorough discussion of the various reductions. The basic algorithm here is $\algoo$ (Definition \ref{def:algoA1}), which iteratively extracts autarkies via the translation until reaching the lean kernel. When combined with cardinality constraints and binary search, indeed $\log_2 n$ oracle calls are sufficient; see Algorithm $\algobs$ (Definition \ref{def:algoAbs}). But these cardinality constraints make the tasks much harder for the SAT oracle. The new algorithm $\algozo$ of this \Schrift{} (Definition \ref{def:alg}) indeed combines the two basic approaches $\algoz, \algoo$, by applying the autarky-resolution duality to the translation and using a more clever choice of ``steering clauses'' to search for autarkies. To better understand this combination of approaches, all four algorithms $\algoz$, $\algoo$, $\algobs$ and $\algozo$, are formulated in a unified way, striving for elegance \emph{and} precision. One feature is, that the input is updated in-place, which not only improves efficiency, but also simplifies the analysis considerably.

\paragraph{Related literature.}

When for $\mc{C}$ (as above) the extended SAT oracle $\oraclezo$ runs in polynomial time, then by \cite[Theorem 11.10.1]{Kullmann2007HandbuchMU} the algorithm $\algoz$ computes a quasi-maximal autarky in polynomial time. The basic applications to 2-CNF, HORN, and the case that every variable occurs at most twice, are reviewed in \cite[Section 11.10.9]{Kullmann2007HandbuchMU}. The other known polytime results regarding computation of the lean kernel use the \emph{deficiency}, as introduced in \cite{FrGe98}, and further studied in \cite{Ku98e}). Here the above algorithm $\algoz$ can not be employed, since crossing out variables can increase this measure (see  \cite[Section 10]{Kullmann2007ClausalFormZI} for a discussion). \cite[Theorem 4.2]{Ku99dKo} shows that the lean kernel is computable in polynomial time for bounded (maximal) deficiency. In \cite{FKS00} the weaker result, that SAT is decidable in polynomial time for bounded maximal deficiency, has been shown, and strengthened later in \cite{Szei2002FixedParam} to fixed-parameter tractability, which is unknown for the computation of the lean kernel. \cite[Theorem 10.3]{Kullmann2007ClausalFormZI} shows that also a maximal autarky can be computed in polynomial time for bounded maximal deficiency, and this for generalised non-boolean clause-sets, connecting to constraint satisfaction.

The connection to the field of hypergraph $2$-colouring, the problem of deciding whether one can colour the vertices of a hypergraph with two colours, such that monochromatic hyperedges are avoided, has been established in \cite{Kullmann2007Balanciert}; see \cite[Section 11.12.2]{Kullmann2007HandbuchMU} and \cite[Subsection 1.6]{KullmannZhao2010Extremal} for overviews. Exploiting the solution of a long-outstanding open problem by \cite{RobertsonSeymourThomas1999GeradeKreise,McCuaig2004PolyasProblem}, the lean kernel is computable in polynomial time by \cite{Kullmann2007Balanciert} for classes of clause-sets, which by \cite[Subsection 1.6]{KullmannZhao2010Extremal}, via the translation of SAT problems into hypergraph $2$-colourability problems, strongly generalises the polytime results (discussed above) for maximal deficiency of clause-sets (partially proven, partially conjectured).

Autarkies have a hidden older history in the field of \emph{Qualitative Matrix Analysis (QMA)}, which yields potential applications of autarky algorithms in economics and elsewhere. QMA was initiated by \cite{Samuelson1947Foundations}, based on the insight that in economics often the magnitude of a quantity is irrelevant, but only the \emph{sign} matters. So \emph{qualitative solvability} of systems of equations and/or inequalities is considered, a special property of such systems, namely that changes of the coefficients, which leave their signs invariant, do not change the signs of the solutions. For a textbook, concentrating on the combinatorial theory, see \cite{BS95}, while a recent overview is \cite{HallLi2007SignPatternMatrices}. The very close connections to autarky theory have been realised in \cite[Section 5]{Ku00f} (motivated by \cite{DD92}), and further expanded in \cite{Kullmann2007Balanciert}; see \cite[Subsection 11.12.1]{Kullmann2007HandbuchMU} for an overview. While preparing this \Schrift{} we came across \cite{KleeLadner1981WeakSAT}, which introduces ``weak satisfiability'', which is \emph{precisely} the existence of a non-trivial autarky. It is shown (\cite[Theorem 5]{KleeLadner1981WeakSAT}), that weak satisfiability is NP-complete; this is the earliest known proof of $\Lean$ being coNP-complete. Apparently these connections to SAT have not been pursued further. The central notions in the early history of QMA were ``$S$-matrix'' and ``$L$-matrix'', which by \cite{Ku00f} are essentially the variable-clause matrices of certain sub-classes of $\Lean$. Unaware of these connections, \cite[Theorem 1.2]{KLM1984} showed directly that recognition of $L$-matrices is coNP-complete. Lean clause-sets correspond to ``$L^+$-matrices'' introduced in \cite{LS1998}, and the decomposition of a clause-set into the lean kernel and the largest autark sub-clause-set now becomes a triangular matrix decomposition into an $L^+$-matrix and the remainder (\cite[Lemma 3.3]{LS1998}).

\paragraph{Applications.}

See \cite{KullmannLynceSilva2005Autarkies} for a general discussion of various redundancy criteria in clause-sets. Identification of maximal autarkies finds application in the analysis of over-constrained systems, for example autark clauses cannot be included in MUSes (minimally unsatisfiable sub-clause-sets) and so, by minimal hitting set duality, cannot be included in MCSes (minimal corrections sets, whose removal leads to a satisfiable clause-set). As discussed above, via the computation of a maximal autarky we can compute basic matrix decompositions of QMA; apparently due to the lack of efficient implementations, at least the related subfield of QMA (which is concerned with NP-hard problems) had yet little practical applications, and the efficient algorithms for computing maximal autarkies via SAT (and extensions) might be a game changer here.

\paragraph{Overview.}

In Section \ref{sec:prelim} we provide all background. Section \ref{sec:oracles} discusses oracles ($\oracle, \oracleo, \oraclez, \oraclezo$), and reviews the first basic algorithm $\algoz$ (Definition \ref{def:A0}), analysed in Lemma \ref{lem:corr0}. Section \ref{sec:trans} introduces the basic translation $F \leadsto t(F)$, where $t(F)$ expresses autarky-search for $F$, and proves various properties. The second basic algorithm $\algoo$ is reviewed in Definition \ref{def:algoA1} and analysed in Lemma \ref{lem:corr1}. Algorithm $\algobs$ is given in Definition \ref{def:algoAbs}, using cardinality constraints (translated into CNF). The use of ``steering clauses'', collected into a set $P$ of positive clauses, is discussed in Subsection \ref{sec:addposcls}, with the main technical result Corollary \ref{cor:readoffaut2}, which shows that variables involved in a resolution refutation of $t(F) \cup P$ can not be part of the largest autarky var-set of $F$. The novel algorithm $\algozo$ finally is introduced in Section \ref{sec:alg}, first using an unspecified $P$ (Definition \ref{def:alg}), and then instantiating this scheme in Theorem \ref{thm:main} to obtain at most $2 \sqrt{n(F)}$ many calls to $\oraclezo$. We conclude in Section \ref{sec:concl} by presenting conjectures and open problems.

\section{Preliminaries}
\label{sec:prelim}

We use $\NN = \set{n \in \ZZ : n \ge 1}$ and $\NNZ = \NN \cup \set{0}$. The powerset of a set $X$ is denoted by $\pot(X)$, while $\pote(X) := \set{X' \in \pot(X) : X' \text{ finite } }$. Maps are sets of ordered pairs, and so for maps $f, g$ the relation $f \sse g$ says, that $f(x) = g(x)$ holds for each $x$ in the domain of $f$, which is contained in the domain of $g$.

We have the set $\Va$ of variables, with $\NN \sse \Va$, and the set $\Lit$ of literals, with $\Va \subset \Lit$. The complementation operation is written $x \in \Lit \mapsto \ol{x} \in \Lit$, and fulfils $\ol{\ol{x}} = x$. On $\NN$ the complementation is arithmetical negation, and thus $\ZZ \sm \set{0} \sse \Lit$. Every literal is either a variable or a complemented variable; forgetting the possible complementation is done by the projection $\var: \Lit \ra \Va$. For $L \sse \Lit$ we use $\ol{L} := \set{\ol{x} : x \in L}$ and $\lit(L) := L \cup \ol{L}$. A clause is a finite set $C \subset \Lit$ of literals with $C \cap \ol{C} = \es$, while a clause-set is a finite set of clauses; the set of all clause-sets is denoted by $\Cls$. The empty clause is denoted by $\bot := \es$, the empty clause-set by $\top := \es \in \Cls$. Furthermore $\Pcls{p} := \set{F \in \Cls : \fa\, C \in F : \abs{C} \le p}$ for $p \in \NNZ$.

For a clause $C$ we define $\var(C) := \set{\var(x) : x \in C}$, while for a clause-set $F$ we define $\var(F) := \bc_{C \in F} \var(C)$. We use the following measures: $n(F) := \abs{\var(F)} \in \NNZ$ is the number of variables, $c(F) := \abs{F} \in \NNZ$ is the number of clauses, $\ell(F) := \sum_{C \in F} \abs{C} \in \NNZ$ is the number of literal occurrences.

A partial assignment is a map $\vp: V \ra \set{0,1}$ for some finite $V \subset \Va$, where we write $\var(\vp) := V$, while the set of all partial assignments is denoted by $\Pass$. A special partial assignment is the empty partial assignment $\epa := \es \in \Pass$. Furthermore we use $\lit(\vp) := \lit(\var(\vp))$, and extend $\vp$ to $\lit(\vp)$ via $\vp(\ol{v}) = 1 - \vp(v)$ for $v \in \var(\vp)$. For $\ve \in \set{0,1}$ we define $\vp^{-1}(\ve) := \set{x \in \lit(\vp) : \vp(x) = \ve}$.

The application $\vp * F \in \Cls$ of $\vp \in \Pass$ to $F \in \Cls$ is defined as $\vp * F := \set{C \sm \vp^{-1}(0) : C \in F \wedge C \cap \vp^{-1}(1) = \es}$. Then $\Sat := \set{F \in \Cls \mb \ex\, \vp \in \Pass : \vp * F = \top}$, and $\Usat := \Cls \sm \Sat$.

The restriction of $F \in \Cls$ to $V \sse \Va$ is defined as $F[V] := \set{C \cap \lit(V) : C \in F} \sm \set{\bot} \in \Cls$, i.e., removal of clauses $C \in F$ with $\var(C) \cap V = \es$, and restriction of the remaining clauses to variables in $V$.

Finally we use $\Cls(V) := \set{F \in \Cls : \var(F) \sse V}$, $\Pass(V) := \set{\vp \in \Pass : \var(\vp) \sse V}$ and $\Tass(V) := \set{\vp \in \Pass : \var(\vp) = V}$ (``total assignments'') for $V \sse \Va$.

Now to autarkies; this \Schrift{} is essentially self-contained, but if more information is desired, see the handbook chapter \cite{Kullmann2007HandbuchMU}. A partial assignment $\vp \in \Pass$ is an \emph{autarky for $F \in \Cls$} iff for all $C \in F$ with $\var(\vp) \cap \var(C) \ne \es$ holds $\vp * \set{C} = \top$ iff $\fa\, C \in F: \vp * \set{C} \in \set{\top,\set{C}}$; the set of all autarkies for $F$ is denoted by $\aut(F) \sse \Pass$. The empty partial assignment $\epa$ is an autarky for every $F \in \Cls$, and in general we call an autarky $\vp$ for $F$ \emph{trivial} if $\var(\vp) \cap \var(F) = \es$. For $\top$ as well as $\set{\bot}$ every partial assignment is a trivial autarky. Note that every satisfying assignment for $F$ is also an autarky for $F$, and it is a trivial autarky iff $F = \top$. Another simple but useful property is that $\vp$ is an autarky for $\bc_{i \in I} F_i$ for a finite family $(F_i)_{i \in I}$ of clause-sets iff $\vp$ is an autarky for all $F_i$, $i \in I$. We also note that $\vp$ is an autarky for $F$ iff $\vp$ is an autarky for $F \cup \set{\bot}$ iff $\vp$ is an autarky for $F \sm \set{\bot}$ (for autarkies the empty clause is invisible). In general it is best to allow that autarkies assign non-occurring variables, but it is also needed to have a notation which disallows this; following \cite[Definition 11.9.1]{Kullmann2007HandbuchMU}:
\begin{definition}\label{def:autf}
  For $F \in \Cls$ let $\bmm{\autf(F)} := \aut(F) \cap \Pass(\var(F))$ (`r'' like ``restricted'' or ``relevant''), while by $\bmm{\var(\autf(F))} := \bc_{\vp \in \autf(F)} \var(\vp)$ we denote the \textbf{largest autarky-var-set}.
\end{definition}

$\Lean \subset \Usat \cup \set{\top}$ is the set of $F \in \Cls$ such that $\autf(F) = \set{\epa}$, while the \emph{lean kernel} of $F \in \Cls$, denoted by $\na(F) \sse F$, is the largest element of $\Lean$ contained in $F$ (it is easy to see that $\Lean$ is closed under finite union). We have $\var(\autf(F)) \cup \var(\na(F)) = \var(F)$ and $\var(\autf(F)) \cap \var(\na(F)) = \es$. See \cite[Subsection 11.8.3]{Kullmann2007HandbuchMU} for various characterisations of the lean kernel.

\begin{definition}\label{def:nval}
  For $F \in \Cls$ let $\bmm{\nva(F)} := \abs{\var(\autf(F))} \in \NNZ$ be the number of variables in the largest autarky-var-set and $\bmm{\nvl(F)} := \abs{\var(\na(F))} \in \NNZ$ be the number of variables in the lean kernel.
\end{definition}
So $n(F) = \nva(F) + \nvl(F)$. On the finite set $\autf(F)$ we have a natural partial order given by inclusion. There is always the smallest element $\epa \in \autf(F)$, while the maximal elements of $\autf(F)$ are called \emph{maximal autarkies} for $F$. For maximal autarkies $\vp, \psi$ holds $\var(\vp) = \var(\psi) = \var(\autf(F))$; here we use that the composition of autarkies is again an autarky, i.e., for autarkies $\vp, \psi$ for $F$ there is an autarky $\theta$ for $F$ with $\vp * (\psi * F) = \psi * (\vp * F) = \theta * F$.

\begin{definition}\label{def:maxaut}
  Let $\bmm{\autmax(F)} \sse \autf(F)$ be the set of maximal autarkies.
\end{definition}

A \emph{quasi-maximal autarky for $F$} is an $\vp \in \autf(F)$ with $\vp * F = \na(F)$. By supplying arbitrary values for the missing variables we obtain efficiently a maximal autarky from a quasi-maximal autarky.

\section{Oracles}
\label{sec:oracles}

The main computational task considered in this \Schrift{} is the computation of some element of $\autmax(F)$ for inputs $F \in \Cls$. Our emphasis is on the number of calls to an ``oracle'', which solves NP-hard problems, while otherwise the computations are in polynomial time. The \textbf{NP (-SAT) oracle} $\bmm{\oracle}: \Cls \ra \set{0,1}$ just maps $F \in \Cls$ to $1$ in case of $F \in \Sat$, and to $0$ otherwise. As we will see in Example \ref{exp:basicalgfindnta}, for deciding leanness, one call suffices. For a \textbf{(standard) SAT oracle} $\bmm{\oracleo}: \Cls \ra \set{0} \cup (\set{1} \times \Pass)$, the SAT solver also returns a satisfying assignment, and then also a non-trivial autarky can be returned in case of non-leanness. As introduced in \cite{Ku01a}, we consider here a strengthened oracle \bmm{\oraclezo}, to return something also for unsatisfiable inputs. Recall that a \emph{tree resolution refutation} for $F \in \Cls$ is a binary tree, where the nodes are labelled with clauses, such that the leaves are labelled by (some) clauses of $F$ (the ``axioms''), while the root is labelled with $\bot$, and such that for each inner node, with children labelled by clauses $C, D$, we have $C \cap \ol{D} = \set{x}$ for some $x \in \Lit$, while the label of that inner node is $(C \sm \set{x}) \cup (D \sm \set{\ol{x}})$.
\begin{definition}\label{def:extsatorac}
  An \textbf{extended SAT oracle} is a map $\bmm{\oraclezo}: \Cls \ra \set{0,1} \times (\pote(\Va) \cup \Pass)$, which for input $F \in \Usat$ returns $(0,\var(F'))$ for some $F' \sse F$, such that there is a tree refutation using as axioms \emph{precisely} $F'$, and for $F \in \Sat$ returns $(1,\vp)$ for some $\vp \in \Pass(\var(F))$ and $\vp * F = \top$. If we don't need the satisfying assignment, then we use $\bmm{\oraclez}: \Cls \ra \set{1} \cup (\set{0} \times \pote(\Va))$.
\end{definition}
In the following we will indicate the type of oracle by using one of $\oraclez, \oracleo, \oraclezo$. See \cite[Subsection 11.10.3]{Kullmann2007HandbuchMU} for a short discussion how to efficiently integrate the computations for $\oraclez, \oraclezo$ into a SAT solver, both look-ahead (\cite{HvM09HBSAT}) and CDCL solvers (\cite{MSLM09HBSAT}). It is important to notice here that we do not need a full resolution refutation, but only the variables involved in it. The above use of \emph{tree} resolution is only a convenient way of stating the condition that all axioms are actually used in the refutation. Furthermore, there is no need for any sort of minimisation of the refutation, as we see by the following lemma.

\begin{lemma}\label{lem:useoracle}
  If for $F \in \Cls$ holds $\oraclez(F) = (0,V)$, then $V \cap \var(\autf(F)) = \es$.
\end{lemma}
\begin{prf}
As shown in \cite[Lemma 3.13]{Ku98e}, for any autarky $\vp \in \aut(F)$ and any clause $C$ touched by $\vp$ there is no tree resolution refutation of $F$ using $C$. \Qed
\end{prf}

So the more clauses are involved in the resolution refutation (i.e., the larger $V$), the more variables we can exclude from the largest autarky-var-set, and thus minimising resolution refutation in general will be counter-productive. One known approach to compute a maximal autarky of $F \in \Cls$, as reviewed in \cite[Subsection 11.10.3]{Kullmann2007HandbuchMU} (especially Theorem 11.10.1 there), is based on the full \emph{autarky-resolution duality} (\cite[Theorem 3.16]{Ku98e}): the variables involved in some autarky of $F$ are altogether, i.e., $\var(\autf(F)) = \var(F) \sm \var(\na(F))$, precisely the variables not usable by some tree resolution refutation of $F$. So the algorithm, called $\algoz(F)$ here, iteratively removes variables not usable in an autarky  and clauses consisting solely of such variables, via Lemma \ref{lem:useoracle}, until a satisfying assignment $\vp$ is found (which must happen eventually), and $\vp$ is then a quasi-maximal (due to autarky-resolution duality):
\begin{definition}\label{def:A0}
  For input $F \in \Cls$, the algorithm \bmm{\algoz(F)}, using oracle $\oraclezo$ and computing a partial assignment $\vp$, performs the following computation:
  \begin{enumerate}
  \item While $\var(F) \ne \es$ do:
    \begin{enumerate}
    \item Compute $\oraclezo(F)$, obtaining $(0,V)$ resp.\ $(1,\vp)$.
    \item In case of $(0,V)$, let $F := F[\var(F) \sm V]$.
    \item In case of $(1,\vp)$, let $F := \top$.
    \end{enumerate}
  \item Return $\vp$.
  \end{enumerate}
\end{definition}

\begin{lemma}[\cite{Ku98e}]\label{lem:corr0}
  For $F \in \Cls$ the algorithm $\algoz(F)$ computes a quasi-maximal autarky for $F$, using at most $\min(\nvl(F)+1, n(F))$ calls of oracle $\oraclezo$.
\end{lemma}

The best case for algorithm $\algoz(F)$ in terms of the number of oracle calls is given for $F \in \Sat$, where just one call suffices. For the worst-case $F \in \Lean$ on the other hand $\algoz(F)$ might use $n(F)$ oracle calls:
\begin{example}\label{exp:ncalls}
  Let $F := \setb{\set{1},\set{-1},\set{2},\set{-2},\dots, \set{n},\set{-n}}$ for $n \in \NNZ$. We have $F \in \Lean$, and each loop iteration will remove exactly one pair $\set{i}, \set{-i}$, until all clauses are removed.
\end{example}

\section{The basic translation}
\label{sec:trans}

We now review the translation $t: \Cls(\Va_0) \ra \Cls$ from \cite{SIMML2014EfficientAutarkies}, called $\Gamma_2$ there, which represents the search for an autarky $\vp$ for $F \in \Cls(\Va_0)$ as a SAT problem $\bmm{t(F)}$; here $\Va_0$ is the set of \emph{primary variables}, while the variables in $\Va \sm \Va_0$ are used as \emph{auxiliary variables}. The translation $t(F)$ uses two types of variables, the primary variables $v \in \var(F)$ themselves, where $v \mapsto 1$ \emph{now} means $v \in \var(\vp)$, and for every $v \in \var(F)$ two auxiliary variables $t(v), t(\ol{v})$, where $t(x) \mapsto 1$ for $x \in \lit(F)$ means $\vp(x) = 1$. In other words, the three possible states of a variable $v \in \var(F)$ w.r.t.\ the partial assignment $\vp$, namely ``unassigned'' ($v \notin \var(\vp)$), ``set true'' ($\vp(v)=1$), ``set false'' ($\vp(v)=0$), are represented by three of the four states of assigned variables $t(v), t(\ol{v})$, namely ``unassigned'' is $t(v), t(\ol{v}) \mapsto 0$, ``set true'' is $t(v) \mapsto 1, t(\ol{v}) \mapsto 0$, and ``set false'' is $t(v) \mapsto 0, t(\ol{v}) \mapsto 1$. The variable $v$ \emph{in the translation} $t(F)$ just acts as an indicator variable, showing whether $v$ is involved in the autarky or not. We have then three types of clauses in $t(F)$: the \emph{autarky clauses} for $C \in F$ and $x \in C$, stating that if $x$ gets false by the autarky, then some other literal of $C$ must get true, plus the \emph{AMO (at-most-one) clauses} for $t(v), t(\ol{v})$ and the \emph{connection} between $v$ and $t(v), t(\ol{v})$. It is useful for argumentation to have the more general form $t_V(F)$, where only $\vp$ with $\var(\vp) \sse V$ are considered:

\begin{definition}\label{def:trans}
  We assume a set $\NN \sse \Va_0 \subset \Va$ of ``primary variables'' together with an injection $t: \lit(\Va_0) \ra \Va$, yielding the ``auxiliary variables'', such that $\Va_0 \cap t(\lit(\Va_0)) = \es$ and $\Va_0 \cup t(\lit(\Va_0)) = \Va$. For $V \sse \Va_0$ let $V' := V \cup t(\lit(V))$. In general we define an equivalence relation on $\Va$, where every equivalence class contains (precisely) three elements, namely $v, t(v), t(\ol{v})$ for $v \in \Va_0$. A set $V \sse \Va$ is \textbf{saturated}, if for $v \in V$ and every equivalent $v'$ holds $v' \in V$. The \textbf{saturation} $V \sse \bmm{V'} \sse \Va$ of $V \sse \Va$ is the saturation under this equivalence relation, i.e., addition of all equivalent variables.

Now the translation $t_V: \Cls(\Va_0) \ra \Cls(V')$ for $V \in \pote(\Va_0)$ has the following clauses for $t_V(F)$:
  \begin{enumerate}
  \item[I] for $C \in F$ and $x \in C$ with $\var(x) \in V$ the \textbf{autarky clause} $\set{\ol{t(\ol{x})}} \cup \set{t(y) : y \in C \sm \set{x}, \var(y) \in V}$ (i.e., $t(\ol{x}) \ra \bigvee_{y \in C \sm \set{x}, \var(y) \in V} t(y)$);
  \item[II] for each $v \in V$ the \textbf{AMO-clause} $\set{\ol{t(v)}, \ol{t(\ol{v})}}$;
  \item[III] for each $v \in V$ the clauses of $v \lra (t(v) \vee t(\ol{v}))$, i.e., the three clauses $\set{\ol{v}, t(v), t(\ol{v})}, \set{\ol{t(v)}, v}, \set{\ol{t(\ol{v})}, v}$ (the \textbf{indicator clauses}).
  \end{enumerate}
  Especially $\bmm{t(F)} := t_{\var(F)}(F)$ for $F \in \Cls(\Va_0)$.
\end{definition}
For $F \in \Cls(\Va_0)$ and $V \in \pote(\Va_0)$ holds $\var(t_V(F)) = V' = V \cup t(\lit(V))$, $V \cap t(\lit(V)) = \es$, and $n(t(F)) = 3 n(F)$, $c(t(F)) = \ell(F) + 4 n(F)$. Due to the four AMO- and indicator-clauses, every satisfying assignment for $t_V(F)$ must be total, that is, for $\vp \in \Pass$ with $\vp * t_V(F) = \top$ holds $\var(t_V(F)) \sse \var(\vp)$.

\begin{example}\label{exp:ncallst}
  For $F = \setb{\set{1},\set{-1},\dots, \set{n},\set{-n}}$ as in Example \ref{exp:ncalls}, we have $2 n$ autarky clauses, which are $\set{\ol{t(i)}}$ for $i \in \tb{-n}{n} \sm \set{0}$.
\end{example}

Partial assignments $\vp$ on the primary variables are translated to assignments on the primary+auxiliary variables via $t_{0,V}(\vp)$ (assigning unassigned variables to $0$ in the translation) and $t(\vp)$ (leaving them unassigned), while the backwards direction goes via via $t^{-1}(\vp)$:
\begin{definition}\label{def:transpass}
  For $V \in \pote(\Va_0)$ we define a translation $\bmm{t_{0,V}}: \Pass(V) \ra \Tass(V')$ for $\vp \in \Pass(V)$ by $t_{0,V}(\vp)(v) = 1 \Lra v \in \var(\vp)$ for $v \in V$, while $t_{0,V}(\vp)(t(x)) = 1 \Lra \var(x) \in \var(\vp) \wedge \vp(x) = 1$ for $x \in \lit(V)$.

  The translation $\bmm{t}: \Pass(\Va_0) \ra \Pass$ for $\vp \in \Pass(\Va_0)$ is the partial assignment, where $\var(t(\vp))$ is the saturation of $\var(\vp)$, while $t(\vp)(v) = 1$ for $v \in \var(\vp)$, and $t(\vp)(t(x)) = 1 \Lra \vp(x) = 1$ for $x \in \lit(\vp)$.

  In the other direction, any partial assignment $\vp \in \Pass$ with $\var(\vp)$ saturated yields a partial assignment $\bmm{t^{-1}(\vp)} \in \Pass(\Va_0)$ with $\var(t^{-1}(\vp)) := \vp^{-1}(1) \cap \Va_0$ and $t^{-1}(\vp)(v) = \vp(t(v))$ for $v \in \var(t^{-1}(\vp))$.
\end{definition}
As already stated, $t_{0,V}(\vp)$ makes explicit which variables are unassigned by $\vp$, namely assigning them with $0$, and thus it needs to know $V$, while $t(\vp)$ just leaves them unassigned. We have $t^{-1}(t_{0,V}(\vp)) = t^{-1}(t(\vp)) = \vp$.

\begin{example}\label{exp:tfsat}
  $t_V(F) \in \Sat$ for $F \in \Cls(\Va_0)$ and $V \in \pote(\Va_0)$, since for $t_{0,V}(\epa) = \pab{v \ra 0 : v \in V} \cup \pab{t(x) \ra 0 : x \in \lit(V)}$ we have $t_{0,V}(\epa) * t_V(F) = \top$.
\end{example}

$t(F)$ does its job, i.e., its solutions represent all the autarkies of $F$:
\begin{lemma}[\cite{SIMML2014EfficientAutarkies}]\label{lem:readoffaut1}
  Consider $F \in \Cls(\Va_0)$ and $V \in \pote(\Va_0)$.
  \begin{enumerate}
  \item\label{lem:readoffaut1a} If $\oracleo(t_V(F)) = (1,\vp)$, then $t^{-1}(\vp) \in \autf(F) \cap \Pass(V)$.
  \item\label{lem:readoffaut1b} $t_{0,V}(\vp) * t_V(F) = \top$ for $\vp \in \autf(F) \cap \Pass(V)$.
  \end{enumerate}
\end{lemma}

Before discussing the usage of $t(F)$, we remark that the variables $\var(F) \sse \var(t(F))$ are used purely for a more convenient discussion, while for a practical application they would be dropped, and the translation called $\Gamma_3$ in \cite{SIMML2014EfficientAutarkies} would be used (except possibly for Algorithm $\algobs$ defined later, which uses cardinality constraints): the variables of $t(F)$ then would be just $t(\lit(F))$, and the clauses would be the autarky- and AMO-clauses (only). In our applications $v \in \var(F)$ occurs in the translations only positively, and would be replaced by the two positive literals $t(v), t(\ol{v})$ (together).

\subsection{Basic usages}
\label{sec:transbasicuse}

\begin{example}\label{exp:basicalgfindnta}
  A simple algorithm for finding a non-trivial autarky for $\var(F) \ne \es$ evaluates $\oracleo(t(F) \cup \set{\var(F)})$. By Lemma \ref{lem:readoffaut1} we get, that if the solver returns $0$, then $F \in \Lean$, while if $(1,\vp)$ is returned, then $t^{-1}(\vp)$ is a non-trivial autarky for $F$ (the non-triviality is guaranteed by the additional clause $\var(F)$).
\end{example}

Algorithm $\algoo(F)$, computing a maximal autarky, iterates the algorithm from Example \ref{exp:basicalgfindnta}; the details are as follows, where we formulate the algorithm in such a way that it has the same basic structure as $\algoz$ (recall Definition \ref{def:A0}) and our novel algorithm $\algozo$ (to be given in Definition \ref{def:alg}):
\begin{definition}\label{def:algoA1}
   For input $F \in \Cls(\Va_0)$ the algorithm \bmm{\algoo(F)}, using oracle $\oracleo$ and computing a partial assignment $\vp$, performs the following computation:
  \begin{enumerate}
  \item $\vp := \epa$, $P := \set{\var(F)}$, $F := t(F)$.
  \item While $\var(P) \ne \es$ do:
    \begin{enumerate}
    \item Compute $\oracleo(F \cup P)$, obtaining $0$ resp.\ $(1,\psi)$.
    \item In case of $0$, let $P := \top$ and $F := \top$.
    \item In case of $(1,\psi)$, let $\psi' := t^{-1}(\psi)$, and update $P := P[\var(P) \sm \var(\psi')]$, $F := t(\psi') * F$, and $\vp := \vp \cup \psi'$.

      In words: obtain the autarky $\psi'$ from $\psi$, remove the variables of $\psi'$ from $P$ and $F$, and add $\psi'$ to the result-autarky $\vp$.
    \end{enumerate}
  \item Return $\vp$.
  \end{enumerate}
\end{definition}

\begin{lemma}\label{lem:corr1}
  For $F \in \Cls(\Va_0)$ the algorithm $\algoo(F)$ computes $\vp \in \autmax(F)$, using at most $\min(\nva(F)+1, n(F))$ calls of oracle $\oracleo$.
\end{lemma}
\begin{prf}
The algorithm always terminates, and moreover for the number $m \ge 0$ of executions of the while-body we have $m \le \min(\nva(F)+1, n(F))$, since in each round $P$ gets reduced by some variables from an autarky (due to the choice of $P$). Let $F_{-1}$ be the input, let $F_0 := t(F_{-1})$, and let $F_i$ for $i = 1,\dots,m$ be the current $F$ after execution of $i$-th iteration; similarly, let $P_0$ be the original value of $P$, and let $P_i$ be the current $P$ after the $i$-th iteration, and let $\vp_0 := \epa$, and let $\vp_i$ be the value of $\vp$ after the $i$-th iteration. Finally, let $V_i$ for $i = 1,\dots,m$ be $\var(P_i)$ in case of $0$ resp.\ the value of $\var(\psi')$ after round $i$, and let $W_0 := \var(F_{-1})$, and let $W_i := W_{i-1} \sm V_i$ for $i = 1,\dots,m$. Inductively we show that $F_i = t_{W_i}(\vp_i * F_{-1})$ for $i \in \tb 0m$, where $\vp_i$ is an autarky for $F_{-1}$ by Lemma \ref{lem:readoffaut1}, Part \ref{lem:readoffaut1a}, and $P_i = P_0[W_i]$ for $i \in \tb 1m$, where $W_m = \es$. Variables only vanish as part of some autarky for $F_{-1}$, and thus $\vp_i \in \autmax(F_{-1}[W_0 \sm W_i])$ for $i \in \tb 0m$. \Qed
\end{prf}

The best case for algorithm $\algoo(F)$ in terms of the number of oracle calls is given for $F \in \Lean$, where just one call suffices. For the worst-case $F \in \Sat$ however, $\algoo(F)$ might use $n(F)$ oracle calls:
\begin{example}\label{exp:algoo}
  Let $F := \set{\set{1}, \dots, \set{n}} \in \Sat$ for $n \in \NNZ$. In the worst case (depending on the answers of $\oracleo$), in each call only one unit-clause $\set{i}$ is removed.
\end{example}

The algorithm realising the currently best number of calls to $\oracleo$ uses SAT-encodings of cardinality constraints (see \cite{RM09HBSAT}); different from the literature, we follow our general scheme and iteratively apply the autarkies found:
\begin{definition}\label{def:algoAbs}
  For input $F \in \Cls(\Va_0)$ the algorithm \bmm{\algobs(F)}, using oracle $\oracleo$ and computing a partial assignment $\vp$, performs the following computation:
  \begin{enumerate}
  \item $\vp := \epa$, $n := n(F)$, $V := \var(F)$, $F := t(F)$ ($n$ is an upper bound on the size of a maximal autarky, $V$ is the set of variables potentially used by it).
  \item While $n \ne 0$ do:
    \begin{enumerate}
    \item $m := \ceil{\frac n2}$; let $G$ be a CNF-representation of the cardinality constraint ``$\,\sum_{v \in V} v \ge m$''; compute $\oracleo(F \cup G)$, obtaining $0$ resp.\ $(1,\psi)$.
    \item In case of $0$, let $n := m-1$.
    \item In case of $(1,\psi)$, let $\psi' := t^{-1}(\psi)$, and update $n := n - n(\psi')$, $V := V \sm \var(\psi')$, $F := t(\psi') * F$, and $\vp := \vp \cup \psi'$.
    \end{enumerate}
  \item Return $\vp$.
  \end{enumerate}
\end{definition}

As it should be obvious by now:
\begin{lemma}\label{lem:corrbs}
  For $F \in \Cls(\Va_0)$ the algorithm $\algobs(F)$ computes $\vp \in \autmax(F)$, using at most $\ceil{\log_2(n(F))}$ calls of oracle $\oracleo$ (for $n(F) > 0$).
\end{lemma}

That the upper bound of Lemma \ref{lem:corrbs} is attained, can be seen again with Example \ref{exp:algoo} (in the worst case). We remark that if we allow calls to Partial MaxSAT (see \cite{LM09HBSAT} for an overview), then just one call is enough (as used in \cite{LiffitonSakallah2008Trimming}), and that without cardinality constraints, namely using $t(F)$ as the hard clauses and $\set{v}$ for $v \in \var(F)$ as the soft clauses. Indeed, as shown in \cite[Proposition 1]{SIMML2014EfficientAutarkies}, this translation has a unique ``minimal correction set'' (MCS), i.e., a unique minimal subset of the soft clauses, whose removal yields a satisfiable clause-set, and so any MCS-solver can be used (just one call).

\subsection{Adding positive ``steering'' clauses}
\label{sec:addposcls}

Generalising the use of $P$ in Algorithm $\algoo$, we consider some positive clause-set $P$ over $\var(F)$ (i.e., $P \sse \pot(\var(F))$), and use $t(F) \cup P \in \Cls$ to gain larger autarkies. Note that the elements of $P$ require variables to be in the autarky, and so in general $P$ should contain several shorter clauses, while for $\algoo$ we just used one full clause (containing all variables). If the oracle then yields unsatisfiability, this is no longer the end of the search (due to the lean kernel been reached), since the clauses of $P$ involved in the refutation might not involve all remaining variables. The extended oracle is now needed to tell us which clauses of $P$ were used. To do so, we first note that autarkies for $F$ yield autarkies for $t(F) \cup P$ (where for a simpler algorithm we allow $P$ to contain variables not in $t(F)$):
\begin{lemma}\label{lem:readoffaut2}
  Consider $F \in \Cls(\Va_0)$ and $P \in \pote(\pote(\Va_0))$. For $\vp \in \autf(F)$ we have $t(\vp) \in \autf(t(F) \cup P)$.
\end{lemma}
\begin{prf}
$t(\vp)$ is an autarky for $P$, since $t(\vp)$ does not set variables from $\var(F)$ to $0$. By Lemma \ref{lem:readoffaut1}, Part \ref{lem:readoffaut1b}, we get that $t_0(\vp)$ is a satisfying assignment for $t(\vp)$; now $t(\vp)$ just unsets all triples $v, t(v), t(\ol{v})$ with $v \notin \var(\vp)$, where $t_0(\vp)$ sets these three variables to $0$. Thus obviously $t(\vp)$ is also an autarky for the AMO-clauses and the indicator clauses. Assume an autarky clause $D$ for $C \in F$ and $x \in C$, touched by $t(\vp)$ but not satisfied. Thus there is $y \in C$ with $\var(x) \notin \var(\vp)$ and $\vp(y) = 0$; since $\vp$ is an autarky, there is $y' \in C$ with $\vp(y') = 1$, whence $t(\vp)(t(y')) = 1$ with $t(y') \in C$, contradicting the assumption. \Qed
\end{prf}

Thus the saturation of the largest autarky-var-set of $F$ is contained in the largest autarky-var-set for $t(F) \cup P$:
\begin{corollary}\label{cor:varautt}
  Consider $F \in \Cls(\Va_0)$ and $P \in \pote(\pote(\Va_0))$. Then the set $\var(\autf(t(F) \cup P))$ is saturated and contains $\var(\autf(F))$.
\end{corollary}
\begin{prf}
It remains to show that $\var(\autf(t(F) \cup P))$ is saturated, and this follows by just considering the AMO-clauses and the indicator clauses: If $v$ is assigned, then also $t(v), t(\ol{v})$ need to be assigned for an autarky, while if one of $t(v), t(\ol{v})$ is assigned, then also $v$ needs to be assigned. \Qed
\end{prf}

Using Lemma \ref{lem:useoracle}, we obtain the main insight, that if the oracle yields $(0,V)$ for $t(F) \cup P$, then none of the elements of $V$ are in the largest autarky-var-set:
\begin{corollary}\label{cor:readoffaut2}
  If for $F \in \Cls(\Va_0)$ and $P \in \pote(\pote(\Va_0))$ the oracle yields $\oraclez(t(F) \cup P) = (0,V)$, then $V' \cap \var(\autf(F)) = \es$ (recall Definition \ref{def:trans} for $V'$).
\end{corollary}

\section{The new algorithm}
\label{sec:alg}

We now present the novel algorithm scheme $\schemezo(F,P)$, combining algorithms $\algoz$ (Definition \ref{def:A0}) and $\algoo$ (Definition \ref{def:algoA1}), which takes as input $F \in \Cls$ and additionally $P \sse \pot(\var(F))$, and computes some autarky $\vp \in \autf(F)$; for our current best generic instantiation we specify $P$ in Theorem \ref{thm:main}, obtaining algorithm $\algozo(F)$.

\begin{definition}\label{def:alg}
  For inputs $F \in \Cls(\Va_0)$ and $P \sse \pot(\var(F))$, the algorithm \bmm{\schemezo(F,P)}, using oracle $\oraclezo$ and computing a partial assignment $\vp$, performs the following computation (using the saturation $V'$ as in Definition \ref{def:trans}):
  \begin{enumerate}
  \item $\vp := \epa$, $F := t(F)$.
  \item While $\var(P) \ne \es$ do:
    \begin{enumerate}
    \item Compute $\oraclezo(F \cup P)$, obtaining $(0,V)$ resp.\ $(1,\psi)$.
    \item In case of $(0,V)$, let $V := V'$, $P := P[\var(P) \sm V]$, $F := F[\var(F) \sm V]$.
    \item In case of $(1,\psi)$, let $\psi' := t^{-1}(\psi)$, and update $P := P[\var(P) \sm \var(\psi')]$, $F := t(\psi') * F$, and $\vp := \vp \cup \psi'$.
    \end{enumerate}
  \item Return $\vp$.
  \end{enumerate}
\end{definition}
While $\bot \in P$ is of no real use, it doesn't cause a problem for the algorithm, and will be removed from $P$ in the first round by the restriction (whether the implicit resolution refutation of $t(F) \cup P$ chooses $\bot$ as the refutation or not).

\begin{lemma}\label{lem:corr01}
  For $F \in \Cls(\Va_0)$ and $P \sse \pot(\var(F))$ the algorithm $\schemezo(F,P)$ computes an autarky $\vp \in \autf(F)$. If $\var(P) = \var(F)$, then $\vp \in \autmax(F)$.
\end{lemma}
\begin{prf}
The proof extends the proof of Lemma \ref{lem:corr1}, by extending the handling of the case $\oraclezo(F \cup P) = (0,V)$. The algorithm always terminates, since in each round $P$ gets reduced. Let $m \ge 0$ be the number of executions of the while-body. Let $F_{-1}$ be the input, let $F_0 := t(F_{-1})$, and let $F_i$ for $i = 1,\dots,m$ be the current $F$ after execution of $i$-th iteration; similarly, let $P_0$ be the input-value of $P$, and let $P_i$ be the current $P$ after the $i$-th iteration, and let $\vp_0 := \epa$, and let $\vp_i$ be the value of $\vp$ after the $i$-th iteration. Finally, let $V_i$ for $i = 1,\dots,m$ be the value of $V$ resp.\ $\var(\psi')$ after round $i$, and let $W_0 := \var(F_{-1})$, and let $W_i := W_{i-1} \sm V_i$ for $i = 1,\dots,m$. Inductively we show that $F_i = t_{W_i}(\vp_i * F_{-1})$ for $i \in \tb 0m$, where $\vp_i$ is an autarky for $F_{-1}$ by Lemma \ref{lem:readoffaut1}, Part \ref{lem:readoffaut1a}, and $P_i = P_0[W_i]$ for $i \in \tb 1m$. Since variables vanish from $P$ only by restriction, we have $V_1 \cup \dots V_m \supseteq \var(P)$, and thus $W_m \sse W_0 \sm \var(P)$. Variables only vanish, if either they are realised as not being element of $\var(\autf(F_{-1}))$ (Corollary \ref{cor:readoffaut2}), or as part of some autarky for $F_{-1}$. So $\vp_i \in \autmax(F_{-1}[W_0 \sm W_i])$ for $i \in \tb 0m$, and if $\var(P) = \var(F_{-1})$, then $\vp_m$ is a maximal autarky for $F_{-1}$. \Qed
\end{prf}

If instead of an unrestricted (maximal) autarky $\vp \in \autf(F)$ we want to compute a (maximal) autarky $\vp \in \autf(F)$ with $\var(\vp) \sse V$ for some given $V \sse \Va$, then we may just replace the input $F$ by $F[V]$ (or we choose $P$ with $\bc P = V$, and restrict the result).

\begin{example}\label{exp:alg1}
  The simplest cases for computing maximal autarkies use (I) $P = \set{\var(F)}$ or (II) $P = \set{\set{v} : v \in \var(F)}$.
  In Case I, we essentially obtain $\algoo$ (Definition \ref{def:algoA1}), and $\schemezo(F,P)$ produces autarkies until the lean kernel is reached, so we only have SAT-answers with one final UNSAT-answer. In Case II, the scheme becomes very similar to $\algoz$ (Definition \ref{def:A0}), and we remove elements of $P$ until we obtain the variables of $\var(\autf(F))$, and so we only have UNSAT-answers with one final SAT answer. If $F \in \Lean$, then in Case I only one call of the oracle is needed (as in Example \ref{exp:basicalgfindnta}), while in Case II, for $F$ as in Example \ref{exp:ncalls} we need $n(F)$ oracle calls. On the other hand, if $F \in \Sat$, then in Case I, for $F$ as in Example \ref{exp:algoo} we need $n(F)$ oracle calls, while in Case II only one call of the oracle is needed.
\end{example}

A more intelligent use of $\schemezo$ employs a better $P$, to mix the SAT- and UNSAT-answers of the oracle.

\begin{lemma}\label{lem:upperbound}
   For $F \in \Cls(\Va_0)$ and $P \sse \pot(\var(F))$ with $P \in \Pcls{p}$ ($p \in \NNZ$), algorithm $\schemezo(F,P)$ uses at most $\min(p,\nva(F)) + \min(c(P),\nvl(F))$ oracle calls.
\end{lemma}
\begin{prf}
Every oracle call removes at least one clause from $P$ (in the unsat-case), since $t_V(F) \in \Sat$, or one variable from all clauses of $P$ (in the sat-case). \Qed
\end{prf}

So we need to minimise the sum of the number of clauses in $P$ and the maximal clause-length, which is achieved by using disjoint clauses of size $\sqrt{n(F)}$; by Lemmas \ref{lem:corr01}, \ref{lem:upperbound} we obtain:

\begin{theorem}\label{thm:main}
  Consider $F \in \Cls(\Va_0)$. Choose $P' \sse \pot(\var(F))$ such that $P'$ is a partitioning of $\var(F)$ (the elements are pairwise disjoint and non-empty, the union is $\var(F)$) with $\fa\, V \in P' : \abs{V} \le \ceil{\sqrt{n(F)}}$ and $c(P') \le \ceil{\sqrt{n(F)}}$.

  Such a partitioning $P'$ can be computed in linear time. Algorithm $\bmm{\algozo(F)} := \schemezo(F,P')$ computes a maximal autarky for $F$, using at most $\min(s,\nva(F)) + \min(s,\nvl(F)) \le 2 s$ calls of $\oraclezo$, where $s := \ceil{\sqrt{n(F)}} \in \NNZ$ .
\end{theorem}

Up to the factor, the upper bound of Theorem \ref{thm:main} is attained:
\begin{example}\label{exp:sqrt}
  For $F$ as in Example \ref{exp:ncallst} as well as $F$ as in Example \ref{exp:algoo} we need now $\ceil{\sqrt{n(F)}}$ oracle calls (in the worst-case).
\end{example}

\section{Conclusion and outlook}
\label{sec:concl}

We reviewed the algorithms $\algoz, \algoo, \algobs$ for computing maximal autarkies, using a unified scheme, and presented the new algorithm $\algozo$. We are employing four different types of oracles: $\oracle$ is the basic oracle, just indicating satisfiability resp.\ unsatisfiability, $\oraclez$ in the unsatisfiable case yields the set of variables used by some resolution refutation, $\oracleo$ in the satisfiable case yields a satisfying assignment, while $\oraclezo$ combines these capabilities. We investigated in some depth the translation $F \leadsto t(F)$, which encodes the autarky search for $F$. The complexities of the four algorithms are summarised as follows (with slight inaccuracies), stating the number and type of oracle calls and the call-instances:
\begin{itemize}
\item $\algoz(F)$: $\nvl(F)$ calls of $\oraclezo$, subinstances of $F$.
\item $\algoo(F)$: $\nva(F)$ calls of $\oracleo$, subinstances of $t(F)$ plus one large positive clause.
\item $\algozo(F)$: $\sqrt{n(F)}$ calls of $\oraclezo$, subinstances of $t(F)$ plus positive clauses.
\item $\algobs(F)$: $\log_2(n(F))$ calls of $\oracleo$, subinstances of $t(F)$ plus one varying cardinality constraint in CNF-representation.
\end{itemize}

\begin{question}\label{que:moreint}
  As we can see from Examples \ref{exp:alg1}, \ref{exp:sqrt}, the choice $P'$ from Theorem \ref{thm:main}, instantiating the scheme $\schemezo$ and yielding $\algozo$, can be improved at least in special cases. Are more intelligent choices of $P$ possible, heuristically, for special classes, or even in general? The optimal choice (hard to compute) is $P := \set{\var(\na(F))} \cup \set{\set{v} : v \in \var(\autf(F))}$, which needs two oracle calls.
\end{question}

\begin{question}\label{que:main}
  We conjecture the number $\Omega(\sqrt{n(F)})$ of oracle calls from Theorem \ref{thm:main} to be optimal in general, but the question here is, how to formalise the restrictions to the input of oracle $\oraclezo$ (so that for example the SAT translations of cardinality constraints are excluded). With these restrictions in place, we also conjecture that when only using oracle $\oracleo$ (as algorithm $\algoo$ does (Definition \ref{def:algoA1})), that then in general $\Omega(n(F))$ many calls are needed.
\end{question}

\begin{question}\label{que:perfdiff}
  How do $\algoz, \algoo, \algozo, \algobs$ compare to each other? Are they pairwise incomparable? Is their oracle usage optimal under suitable constraints?
\end{question}

\begin{question}\label{que:leancomp}
  In this \Schrift{} we concentrated on the hardest functional task: What about the complexity of the computation of the lean kernel, when using oracles $\oracle, \oraclez, \oracleo, \oraclezo$ ? Do we need less calls than for computing maximal autarkies?
\end{question}

Only one precise conjecture on lower bounds for the computation of maximal autarkies seems possible currently:
\begin{conjecture}\label{que:storac}
  The computation of a maximal autarky for input $F \in \Cls$, when using a SAT oracle $\oracleo$, in general needs $\Omega(\log_2(n(F)))$ many calls; possibly one can even show that for every (deterministic) algorithm there exists an instance needing at least $\log_2(n(F))$ many calls.
\end{conjecture}

Finally we remark that for the considerations of this \Schrift{} more fine-grained complexity notions for function classes and their oracle usage are needed. Function classes just using NP-oracles (only returning yes/no) have been studied starting with \cite{Krentel1988Optimisation}, while a systematic study of ``function oracles'' has been started in \cite{MarquesSilvaJanota2014QueryComplexity}, using ``witness oracles''; we note that $\oraclez, \oraclezo$ are not such witness oracles (we can not easily check the returned var-sets).

\bibliographystyle{plainurl}

\begin{thebibliography}{10}

\bibitem{2008HandbuchSAT}
Armin Biere, Marijn~J.H. Heule, Hans van Maaren, and Toby Walsh, editors.
\newblock {\em Handbook of Satisfiability}, volume 185 of {\em Frontiers in
  Artificial Intelligence and Applications}.
\newblock IOS Press, February 2009.

\bibitem{BS95}
Richard~A. Brualdi and Bryan~L. Shader.
\newblock {\em Matrices of sign-solvable linear systems}, volume 116 of {\em
  Cambridge Tracts in Mathematics}.
\newblock Cambridge University Press, 1995.
\newblock ISBN 0-521-48296-8.
\newblock \href {http://dx.doi.org/10.1017/CBO9780511574733}
  {\path{doi:10.1017/CBO9780511574733}}.

\bibitem{DD92}
Gennady Davydov and Inna Davydova.
\newblock Tautologies and positive solvability of linear homogeneous systems.
\newblock {\em Annals of Pure and Applied Logic}, 57(1):27--43, May 1992.
\newblock \href {http://dx.doi.org/10.1016/0168-0072(92)90060-D}
  {\path{doi:10.1016/0168-0072(92)90060-D}}.

\bibitem{EIS76}
S.~Even, A.~Itai, and A.~Shamir.
\newblock On the complexity of timetable and multicommodity flow problems.
\newblock {\em SIAM Journal Computing}, 5(4):691--703, 1976.
\newblock \href {http://dx.doi.org/10.1137/0205048}
  {\path{doi:10.1137/0205048}}.

\bibitem{FKS00}
Herbert Fleischner, Oliver Kullmann, and Stefan Szeider.
\newblock Polynomial--time recognition of minimal unsatisfiable formulas with
  fixed clause--variable difference.
\newblock {\em Theoretical Computer Science}, 289(1):503--516, November 2002.
\newblock \href {http://dx.doi.org/10.1016/S0304-3975(01)00337-1}
  {\path{doi:10.1016/S0304-3975(01)00337-1}}.

\bibitem{FrGe98}
John Franco and Allen~Van Gelder.
\newblock A perspective on certain polynomial-time solvable classes of
  satisfiability.
\newblock {\em Discrete Applied Mathematics}, 125(2-3):177--214, 2003.
\newblock \href {http://dx.doi.org/10.1016/S0166-218X(01)00358-4}
  {\path{doi:10.1016/S0166-218X(01)00358-4}}.

\bibitem{HallLi2007SignPatternMatrices}
Frank~J. Hall and Zhongshan Li.
\newblock Sign pattern matrices.
\newblock In Leslie Hogben, editor, {\em Handbook of Linear Algebra}, Discrete
  Mathematics and Its Applications, pages 33(1)--33(21). Chapman \& Hall/CRC,
  2007.
\newblock ISBN 1-58488-510-6.
\newblock \href {http://dx.doi.org/10.1201/9781420010572.ch33}
  {\path{doi:10.1201/9781420010572.ch33}}.

\bibitem{HvM09HBSAT}
Marijn J.~H. Heule and Hans van Maaren.
\newblock Look-ahead based {SAT} solvers.
\newblock In Biere et~al. \cite{2008HandbuchSAT}, chapter~5, pages 155--184.
\newblock \href {http://dx.doi.org/10.3233/978-1-58603-929-5-155}
  {\path{doi:10.3233/978-1-58603-929-5-155}}.

\bibitem{KleeLadner1981WeakSAT}
Victor Klee and Richard Ladner.
\newblock Qualitative matrices: Strong sign-solvability and weak
  satisfiability.
\newblock In Harvey~J. Greenberg and John~S. Maybee, editors, {\em
  Computer-Assisted Analysis and Model Simplification}, pages 293--320, 1981.
\newblock Proceedings of the First Symposium on Computer-Assisted Analysis and
  Model Simplification, University of Colorado, Boulder, Colorado, March 28,
  1980.
\newblock \href {http://dx.doi.org/10.1016/B978-0-12-299680-1.50022-7}
  {\path{doi:10.1016/B978-0-12-299680-1.50022-7}}.

\bibitem{KLM1984}
Victor Klee, Richard Ladner, and Rachel Manber.
\newblock Signsolvability revisited.
\newblock {\em Linear Algebra and its Applications}, 59:131--157, June 1984.
\newblock \href {http://dx.doi.org/10.1016/0024-3795(84)90164-2}
  {\path{doi:10.1016/0024-3795(84)90164-2}}.

\bibitem{Kullmann2007HandbuchMU}
Hans {Kleine B{\"{u}}ning} and Oliver Kullmann.
\newblock Minimal unsatisfiability and autarkies.
\newblock In Biere et~al. \cite{2008HandbuchSAT}, chapter~11, pages 339--401.
\newblock \href {http://dx.doi.org/10.3233/978-1-58603-929-5-339}
  {\path{doi:10.3233/978-1-58603-929-5-339}}.

\bibitem{Krentel1988Optimisation}
Mark~W. Krentel.
\newblock The complexity of optimization problems.
\newblock {\em Journal of Computer and System Sciences}, 36(3):490--509, June
  1988.
\newblock \href {http://dx.doi.org/10.1016/0022-0000(88)90039-6}
  {\path{doi:10.1016/0022-0000(88)90039-6}}.

\bibitem{Ku99dKo}
Oliver Kullmann.
\newblock An application of matroid theory to the {SAT} problem.
\newblock In {\em Proceedings of the 15th Annual IEEE Conference on
  Computational Complexity}, pages 116--124, July 2000.
\newblock \href {http://dx.doi.org/10.1109/CCC.2000.856741}
  {\path{doi:10.1109/CCC.2000.856741}}.

\bibitem{Ku98e}
Oliver Kullmann.
\newblock Investigations on autark assignments.
\newblock {\em Discrete Applied Mathematics}, 107:99--137, 2000.
\newblock \href {http://dx.doi.org/10.1016/S0166-218X(00)00262-6}
  {\path{doi:10.1016/S0166-218X(00)00262-6}}.

\bibitem{Ku01a}
Oliver Kullmann.
\newblock On the use of autarkies for satisfiability decision.
\newblock In Henry Kautz and Bart Selman, editors, {\em LICS 2001 Workshop on
  Theory and Applications of Satisfiability Testing (SAT 2001)}, volume~9 of
  {\em Electronic Notes in Discrete Mathematics (ENDM)}, pages 231--253.
  Elsevier Science, June 2001.
\newblock \href {http://dx.doi.org/10.1016/S1571-0653(04)00325-7}
  {\path{doi:10.1016/S1571-0653(04)00325-7}}.

\bibitem{Ku00f}
Oliver Kullmann.
\newblock Lean clause-sets: Generalizations of minimally unsatisfiable
  clause-sets.
\newblock {\em Discrete Applied Mathematics}, 130:209--249, 2003.
\newblock \href {http://dx.doi.org/10.1016/S0166-218X(02)00406-7}
  {\path{doi:10.1016/S0166-218X(02)00406-7}}.

\bibitem{Kullmann2007Balanciert}
Oliver Kullmann.
\newblock Polynomial time {SAT} decision for complementation-invariant
  clause-sets, and sign-non-singular matrices.
\newblock In Joao Marques-Silva and Karem~A. Sakallah, editors, {\em Theory and
  Applications of Satisfiability Testing - SAT 2007}, volume 4501 of {\em
  Lecture Notes in Computer Science}, pages 314--327. Springer, 2007.
\newblock \href {http://dx.doi.org/10.1007/978-3-540-72788-0_30}
  {\path{doi:10.1007/978-3-540-72788-0_30}}.

\bibitem{Kullmann2007ClausalFormZI}
Oliver Kullmann.
\newblock Constraint satisfaction problems in clausal form {I}: Autarkies and
  deficiency.
\newblock {\em Fundamenta Informaticae}, 109(1):27--81, 2011.
\newblock \href {http://dx.doi.org/10.3233/FI-2011-428}
  {\path{doi:10.3233/FI-2011-428}}.

\bibitem{Kullmann2007ClausalFormZII}
Oliver Kullmann.
\newblock Constraint satisfaction problems in clausal form {II}: Minimal
  unsatisfiability and conflict structure.
\newblock {\em Fundamenta Informaticae}, 109(1):83--119, 2011.
\newblock \href {http://dx.doi.org/10.3233/FI-2011-429}
  {\path{doi:10.3233/FI-2011-429}}.

\bibitem{KullmannLynceSilva2005Autarkies}
Oliver Kullmann, In{\^{e}}s Lynce, and Jo{\~{a}}o Marques-Silva.
\newblock Categorisation of clauses in conjunctive normal forms: Minimally
  unsatisfiable sub-clause-sets and the lean kernel.
\newblock In Armin Biere and Carla~P. Gomes, editors, {\em Theory and
  Applications of Satisfiability Testing - SAT 2006}, volume 4121 of {\em
  Lecture Notes in Computer Science}, pages 22--35. Springer, 2006.
\newblock \href {http://dx.doi.org/10.1007/11814948_4}
  {\path{doi:10.1007/11814948_4}}.

\bibitem{KullmannZhao2011Bounds}
Oliver Kullmann and Xishun Zhao.
\newblock On variables with few occurrences in conjunctive normal forms.
\newblock In Laurent Simon and Karem Sakallah, editors, {\em Theory and
  Applications of Satisfiability Testing - SAT 2011}, volume 6695 of {\em
  Lecture Notes in Computer Science}, pages 33--46. Springer, 2011.
\newblock \href {http://dx.doi.org/10.1007/978-3-642-21581-0_5}
  {\path{doi:10.1007/978-3-642-21581-0_5}}.

\bibitem{KullmannZhao2010Extremal}
Oliver Kullmann and Xishun Zhao.
\newblock Bounds for variables with few occurrences in conjunctive normal
  forms.
\newblock Technical Report arXiv:1408.0629v3 [math.CO], arXiv, November 2014.
\newblock Available from: \url{http://arxiv.org/abs/1408.0629}.

\bibitem{LS1998}
Gwang-Yeon Lee and Bryan~L. Shader.
\newblock Sign-consistency and solvability of constrained linear systems.
\newblock {\em The Electronic Journal of Linear Algebra}, 4:1--18, August 1998.
\newblock Available from:
  \url{http://emis.matem.unam.mx/journals/ELA/ela-articles/4.html}.

\bibitem{LM09HBSAT}
Chu~Min Li and Felip Many{\`a}.
\newblock {MaxSAT}, hard and soft constraints.
\newblock In Biere et~al. \cite{2008HandbuchSAT}, chapter~19, pages 613--631.
\newblock \href {http://dx.doi.org/10.3233/978-1-58603-929-5-613}
  {\path{doi:10.3233/978-1-58603-929-5-613}}.

\bibitem{LiffitonSakallah2008Trimming}
Mark Liffiton and Karem Sakallah.
\newblock Searching for autarkies to trim unsatisfiable clause sets.
\newblock In Hans {Kleine B{\"{u}}ning} and Xishun Zhao, editors, {\em Theory
  and Applications of Satisfiability Testing - SAT 2008}, volume 4996 of {\em
  Lecture Notes in Computer Science}, pages 182--195. Springer, 2008.
\newblock \href {http://dx.doi.org/10.1007/978-3-540-79719-7_18}
  {\path{doi:10.1007/978-3-540-79719-7_18}}.

\bibitem{SIMML2014EfficientAutarkies}
Joao Marques-Silva, Alexey Ignatiev, Antonio Morgado, Vasco Manquinho, and Ines
  Lynce.
\newblock Efficient autarkies.
\newblock In Torsten Schaub, Gerhard Friedrich, and Barry O'Sullivan, editors,
  {\em 21st European Conference on Artificial Intelligence (ECAI 2014)}, volume
  263 of {\em Frontiers in Artificial Intelligence and Applications}, pages
  603--608. IOS Press, 2014.
\newblock \href {http://dx.doi.org/10.3233/978-1-61499-419-0-603}
  {\path{doi:10.3233/978-1-61499-419-0-603}}.

\bibitem{MarquesSilvaJanota2014QueryComplexity}
Joao Marques-Silva and Mikol{\'{a}}{\v{s}} Janota.
\newblock On the query complexity of selecting few minimal sets.
\newblock Technical Report TR14-031, Electronic Colloquium on Computational
  Complexity (ECCC), March 2014.
\newblock Available from: \url{http://eccc.hpi-web.de/report/2014/031/}.

\bibitem{MSLM09HBSAT}
Joao~P. Marques-Silva, Ines Lynce, and Sharad Malik.
\newblock Conflict-driven clause learning {SAT} solvers.
\newblock In Biere et~al. \cite{2008HandbuchSAT}, chapter~4, pages 131--153.
\newblock \href {http://dx.doi.org/10.3233/978-1-58603-929-5-131}
  {\path{doi:10.3233/978-1-58603-929-5-131}}.

\bibitem{McCuaig2004PolyasProblem}
William McCuaig.
\newblock P{\'{o}}lya's permanent problem.
\newblock {\em The Electronic Journal of Combinatorics}, 11, 2004.
\newblock \#R79, 83 pages.
\newblock Available from:
  \url{http://www.combinatorics.org/ojs/index.php/eljc/article/view/v11i1r79}.

\bibitem{MoSp85}
B.~Monien and Ewald Speckenmeyer.
\newblock Solving satisfiability in less than $2^n$ steps.
\newblock {\em Discrete Applied Mathematics}, 10(3):287--295, March 1985.
\newblock \href {http://dx.doi.org/10.1016/0166-218X(85)90050-2}
  {\path{doi:10.1016/0166-218X(85)90050-2}}.

\bibitem{Ok98}
Fumiaki Okushi.
\newblock Parallel cooperative propositional theorem proving.
\newblock {\em Annals of Mathematics and Artificial Intelligence},
  26(1-4):59--85, 1999.
\newblock \href {http://dx.doi.org/10.1023/A:1018946526109}
  {\path{doi:10.1023/A:1018946526109}}.

\bibitem{PW88}
Christos~H. Papadimitriou and David Wolfe.
\newblock The complexity of facets resolved.
\newblock {\em Journal of Computer and System Sciences}, 37(1):2--13, August
  1988.
\newblock \href {http://dx.doi.org/10.1016/0022-0000(88)90042-6}
  {\path{doi:10.1016/0022-0000(88)90042-6}}.

\bibitem{RobertsonSeymourThomas1999GeradeKreise}
Neil Robertson, Paul~D. Seymour, and Robin Thomas.
\newblock Permanents, {P}faffian orientations, and even directed circuits.
\newblock {\em Annals of Mathematics}, 150(3):929--975, 1999.
\newblock \href {http://dx.doi.org/10.2307/121059} {\path{doi:10.2307/121059}}.

\bibitem{RM09HBSAT}
Olivier Roussel and Vasco Manquinho.
\newblock Pseudo-boolean and cardinality constraints.
\newblock In Biere et~al. \cite{2008HandbuchSAT}, chapter~22, pages 695--733.
\newblock \href {http://dx.doi.org/10.3233/978-1-58603-929-5-695}
  {\path{doi:10.3233/978-1-58603-929-5-695}}.

\bibitem{Samuelson1947Foundations}
Paul~A. Samuelson.
\newblock {\em Foundations of Economic Analysis}.
\newblock Harvard University Press, 1947.

\bibitem{Szei2002FixedParam}
Stefan Szeider.
\newblock Minimal unsatisfiable formulas with bounded clause-variable
  difference are fixed-parameter tractable.
\newblock {\em Journal of Computer and System Sciences}, 69(4):656--674,
  December 2004.
\newblock \href {http://dx.doi.org/10.1016/j.jcss.2004.04.009}
  {\path{doi:10.1016/j.jcss.2004.04.009}}.

\end{thebibliography}

\newcommand{\noopsort}[1]{}

\end{document}